\documentclass[twocolumn, arguments]{aastex631}

\shorttitle{X10 flares in solar cycles}
\shortauthors{Tan et al.}
\graphicspath{{./}{figures/}}
\usepackage{xcolor}

\begin{document}

\title{The occurrence of powerful flares stronger than X10 class in Solar Cycles}

\email{bltan@nao.cas.cn}

\author{Baolin Tan$^{1, 2}$, Yin Zhang$^{1}$, Jing Huang}

\affil{National Astronomical Observatories of Chinese Academy of Sciences, Datun Road 20A, Chaoyang District, Beijing 100012, China; bltan@nao.cas.cn}

\affil{School of Astronomy and Space Sciences, University of Chinese Academy of Sciences, Beijing 100049, China.}

\author{Kaifan Ji}

\affil{Yunnan Observatories, Chinese Academy of Sciences, P.O. Box 110, Kunming 650011, China.}
\affil{Yunnan Key Laboratory of Solar Physics and Space Science, Kunming 650216, China.}

\begin{abstract}

Solar flares stronger than X10 (S-flares, $>$X10) are the highest class flares which significantly impact on the Sun's evolution and space weather. Based on observations of Geostationary Orbiting Environmental Satellites (GOES) at soft X-ray (SXR) wavelength and the daily sunspot numbers (DSNs) since 1975, we obtained some interesting and heuristic conclusions: (1) Both S-flares and the more powerful extremely strong flares (ES-flares, $>$X14.3) mostly occur in the late phases of solar cycles and low-latitude regions on the solar disk; (2) Similar to X-class flares, the occurrence of S-flares in each solar cycle is somewhat random, but the occurrence of ES-flares seems to be dominated by the mean DSN ($V_{m}$) and its root-mean-square deviation during the valley phase ($V_{d}$) before the cycle: the ES-flare number is strongly correlated with $V_{d}$, and the occurrence time of the first ES-flare is anti-correlated with $V_{d}$ and $V_{m}$. These facts indicate that the higher the $V_{m}$ and $V_{d}$, the stronger the solar cycle, the more the ES-flares and the earlier they occurred. We proposed that the Sun may have a low-latitude active zone (LAZ), and most ES-flares are generated from the interaction between LAZ and the newly emerging active regions. The correlations and the linear regression functions may provide an useful method to predict the occurrence of ES-flares in an upcoming solar cycle, which derives that solar cycle 25 will have about $2\pm1$ ES-flares after the spring of 2027.

\end{abstract}


\section{Introduction}

Solar eruptions refer to the rapid and violent magnetic energy release in the low atmosphere of the Sun, including solar flares, coronal mass ejections (CMEs), and various plasma jets. Among them, solar flares are the most spectacular events, CMEs and jets are more or less physically linked to flares (Harrison 1995). Generally, solar flares are classified into A, B, C, M, and X-class (Bai \& Sturrock 1989), which denote the order of peak flux of soft X-ray (SXR) emission ($F_{sxr}$) observed by Geostationary Orbiting Environmental Satellites (GOES) at wavelength of 1 - 8 {\AA} on a logarithmic scale: $A = 10^{-8}$, $B = 10^{-7}$, $C = 10^{-6}$, $M = 10^{-5}$, and $X = 10^{-4}$ $W\cdot m^{-2}$, respectively, and the subdivided class is added an additional digit, such as M5 means a solar flare with SXR emission peak flux of $5.0\times 10^{-5}$ $W\cdot m^{-2}$. When $F_{sxr}>10^{-3}$ $W\cdot m^{-2}$, that means the flare is stronger than X10, and can be denoted as S-flare. Thousands of flares occurred in each solar cycle (SC). For example, SC 23 had 6 S-flares, 167 X-class flares, 1444 M-class flares, and 12995 C-class flares. Although, according to the definition of super flares (Schaefer et al. 2000, Maehara et al. 2012, Shibayama et al. 2013, Notsu et al. 2019), even the strongest solar flares recorded in history (for example, Carrington event on 1859 September 1 with class of nearly X80, Hayakawa et al. 2023) still did not meet the criterion of super flares (Emslie et al. 2012, Cliver et al. 2022). However, S-flares are the most powerful eruptions occurred in the whole solar system and have the greatest impact on the space environment. This work attempts to identify certain patterns of S-flares from long-term observational data on SXR/GOES and the daily sunspot number (DSN) since 1975, which may be beneficial for predicting solar activity and studying the long-term evolution of the Sun.

Many people predicted that SC 25 should be stronger than SC 24 (Luo \& Tan 2024, Jiang et al. 2023, Guo et al. 2021, Pesnell \& Schatten 2018, etc.). Since the start of SC 25 from August 11, 2019, it has gone through more than 5.31 yr (up to November 30, 2024) and has actually entered its peak phase. So far, the strongest flare was an X9.0 flare occurred on October 3, 2024, and no S-flare occurred yet. We can't help but ask, will there be S-flare in SC 25? how many S-flares will occur? when will they occur? will there be more powerful flares during the SC 25? These questions are very interesting and important that has received widespread attentions and is extremely challenging for us (Schrijver 2007, Bloomfield et al. 2012, Khlystov, 2014, Upton \& Hathaway 2023).

Solar observations have become increasingly complicated: expensive multiple instruments, multiple frequency bands, high-resolutions, big data, and various large-scale and complicated analysis techniques. However, we are still unable to effectively predict solar eruptions. S-flare and the more powerful flares should be the most spectacular large-scale explosions on the Sun. They should likely to respond to the simplest solar activity parameters, such as DSN. If we could find some relatively simple indicator from the DSN records of the past 4 solar cycles to predict the occurrence of S-flares or even more powerful flares in the upcoming solar cycle in advance, it will be greatly helpful, valuable, and operable. Section 2 introduces the main data used in this study, as well as the definition and extraction methods of relevant parameters. Section 3 presents the main results, linear regression functions, predicting results for SC 25, and the possible physical explanation. The conclusions are summarized in Section 4.

\section{Data and parameters extraction}

\subsection{Data source}

Here, we selected two kinds of long-term recorded observational data for our study.

(1) The relative daily sunspot numbers (DSN). It is one of the global index of solar magnetic activities. The DSN data comes from the World Data Center SILSO where has adopted an open data policy, the data are corrected and assembled by standard methods (Clette and Lefevre 2016), and can be freely downloaded from the network(WDC-SILSO, Royal Observatory of Belgium, Brussels. Website at: https://www.sidc.be/index.php/SILSO/datafiles). Here, we directly obtain DSN records since 1818 January 1. Detailed information about the various diagnostics and corrections of DSN data can be found in Li et al. (2002), and Clette et al. (2014).

(2) The flare list observed at SXR wavelengths by GOES satellites. It is obtained from the GOES/SXR observation at the 1.0 - 8.0 \AA~ and 0.5 - 4.0 \AA~ wavelengths since 1975. GOES is a series of satellites, including GOES-1 to GOES-15 satellites from 1975 to 2017, and GOES-R series satellites since 2017, including the current GOES-16, GOES-17, and GOES-18 at present, and GOES-19 in the future. GEOS SXR observations provide a list of flares, including the location of flaring active region on the solar disk. Hudson et al. (2024) found that SXR flares recorded by the old GOES series (GOES-1 to GOES-15) should be multiplied a correction factor 1.43 to be consistent with GOES-R series (GOES-16 to GOES-19). Additionally, saturation also should be corrected for the strong flares (X-class and beyond). For example, after correcting the saturation, the previous listed X15 flare on 1978 July 11 is now X45.9, which is the strongest flare since 1975. The original X9.3 flare on 2017 September 6 is actually classified X14.8, which now becomes an S-flare. Totally, there are 37 S-flares during 1975-2024, which are marked in Figure 1 and listed in Table 1.

Figure 1 presents the distribution of S-flares on the profile of DSN during from 1975 to 2024. There is an overlapped smoothed curve (red) of DSN which is designed to show clearly the valley and peak times of each cycle. The width of smoothed window is chosen to ensure that each cycle has exactly single valley and single peak. The purpose of doing this is to accurately pinpoint the start, peak, and end times of each cycle, and to facilitate the phase of the cycle. Here, we obtained 5 valley points marked as V21, V22, V23, V24, and V25 (the black dot-dashed vertical lines). They show the starts of SC 21, 22, 23, 24, and 25, and also show the end of SC 20, 21, 22, 23, and 24. The four red dot-dashed vertical lines show the peak points of SC 21, 22, 23, and 24, respectively. The pluses (+) marked S-flares.

\begin{figure*} 
\begin{center}
   \includegraphics[width=15 cm]{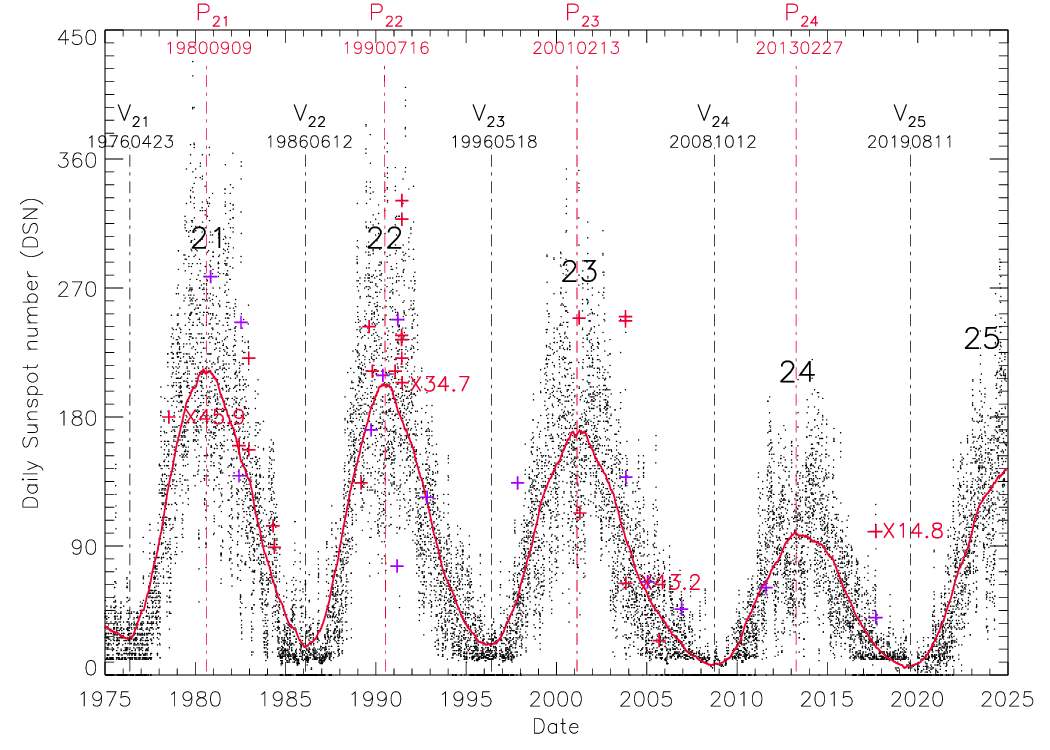}
\caption{The distribution of the very large flares stronger than X10 (S-flare) during solar cycle 21-25. The black dots are the daily sunspot numbers (DSN) with a smoothed curve (red). The purple pluses (+) mark the S-flares with class of X10 - X14.3 and the red pluses (+) mark the ES-flares ($>X14.3$). The vertical black and red dash-dotted lines mark the valley and peak points of each cycle, respectively. Here, we also highlight the strongest flares for each solar cycle.}
\end{center}
\end{figure*}

Figure 1 shows that S-flares do not directly respond to the very high DSN, some S-flares occurred even with very small DSN. A similar conclusion for X-class flares was reached by Hudson (2007). Therefore, we cannot simply use DSN to predict an upcoming S-flare.

\subsection{Parameter Definitions}

Our question about S-flares can be reflected in the following three parameters:

(1) How many S-flares occurred in a solar cycle? $N_{s}$.

(2) When does the first S-flare occur during a solar cycle? $t_{s1}$, which is defined as the time difference between the start of a cycle and the occurrence of the first S-flare.

(3) The mean time of all S-flares in a solar cycle, $t_{sm}$. The time of an S-flare $t$ is defined as the time difference between the start of the cycle and the onset of the S-flare. $t_{sm}$ is defined as the time average of all S-flares occurred in a cycle.

Tan (2019) reported that the mean DSN of valley phase was correlated with the magnitude of the forthcoming cycle. Therefore, it is possible to apply the parameters of valley phase to predict the occurrence of S-flares in advance before the start of cycles. Here, we define the following parameters to describe the valley phase of solar cycle:

(1) The mean DSN during valley phase, $V_{m}$, it reflects the valley level before the start of a cycle.

(2) The root-mean-square deviation of DSN during valley phase, $V_{d}$, it reflects the degree of variation of DSN around the valley phase.

In order to study the corresponding physical mechanism, we also extracted the following parameters to describe the cycle:

(1) Period of solar cycle, $P_{c}$, defined as the time interval between the start and end of a cycle.

(2) Length of ascending phase of solar cycle, $L_{a}$, defined as the time interval between the start and peak of a cycle.

(3) Length of descending phase of solar cycle, $L_{d}$, defined as the time interval between the peak and end of a cycle.

Obviously, we have $P_{c}=L_{a}+L_{d}$.

(4) The maximum DSN of a solar cycle, $M_{sn}$.

(5) The mean DSN during the peak phase, $P_{m}$, it reflects the solar activity level of a cycle.

(6) The root-mean-square deviation of DSN during the peak phase, $P_{d}$, it reflects the degree of variation of DSN around the peak phase.

All above parameters for SC 21-24 can be extracted from Figure 1 and listed in Table 2. The width of valley or peak phases is defined as a period during which DSN changes relatively flat before and after the valley or peak points. Here we hope to remove the influence of the last descending phase and the next ascending phase. Based on careful comparisons, we found that when the half width of valley or peak phase was set as 150 days  (totaly 300 days), the data is relatively flat; beyond this width, the trend of the data changes in an arc, which may contain influence of the last ascending phase or the next descending phase.

\section{Results, Predictions and the Related Mechanism}

\subsection{The correlations between the occurrence of S-flares and the parameters of solar cycles}

Figure 1 shows that S-flares do not strongly depend on DSN. In SC 23, the strongest flare (X43.2, on 2003 November 4) corresponds to a DSN of 64, while the maximum DSNs of this cycle is high up to 353. In fact, when DSN reaches to the maximum of a solar cycle, there is always no powerful flare occurred. For example, DSN reached to maximum of SC 24 on 2014 February 27, but that day only produced 12 C-class flares, no more powerful flare. The other cycles also have the similar trend. These facts indicate that there is no obvious correlation between DSN and the occurrence of S-flares. It is difficult to predict the occurrence of S-flares simply from DSN.

Table 1 listed the parameters of 37 S-flares, including their date, the corresponding DSN, active region and its maximum area and position on the solar disk, and the time of the occurrence after the start of the corresponding solar cycle.

\begin{deluxetable}{ccccccccccccccccc}
\tablecolumns{16} \tabletypesize{\scriptsize} \tablewidth{0pc}
\tablecaption{List of the solar S-class flares during 1976-2024.  \label{tbl-1}}
\tablehead{\colhead{No.} &\colhead{Date} &\colhead{DSN} &\colhead{Class} &\colhead{AR} &\colhead{$S_{AR}$} &\colhead{Location} &\colhead{$t$(yr)} }
  \startdata
1  & 19780711    &  180          &    X45.9     & 1203   & 1370   &  N20E46  & 2.22  \\
2  & 19801106    &  278          &    X12.8     & 2776   & 1440   &  S12E74  & 4.54  \\
3  & 19820603    &  139          &    X12.8     & 3763   & 1250   &  S09E72  & 6.11  \\
4  & 19820606    &  160          &    X14.7     & 3763   & 1250   &  S09E25  & 6.12  \\
5  & 19820712    &  246          &    X10.1     & 3804   & 2870   &  N11E37  & 6.22  \\
6  & 19821215    &  221          &    X19.0     & 4026   & 640    &  S10E24  & 6.64  \\
7  & 19821217    &  157          &    X14.7     & 4025   & --     &  S08W21  & 6.65  \\
8  & 19840424    &  104          &    X19.7     & 4474   & 2590   &  S12E43  & 8.00  \\
9  & 19840520    &  89           &    X14.8     & 4492   & 670    &  S09E52  & 8.07  \\
10 & 19890306    &  134          &    X19.3     & 5395   & 3570   &  N35E69  & 3.14  \\
11 & 19890816    &  243          &    X28.0     & 5629   & 1320   &  S18W84  & 3.58  \\
12 & 19890929    &  171          &    X14.2     & 5698   & 1250   &  S20W90  & 3.70  \\
13 & 19891019    &  212          &    X23.1     & 5747   & 1160   &  S27E10  & 3.76  \\
14 & 19900524    &  209          &    X14.1     & 6063   &  940   &  N33W78  & 4.35  \\
15 & 19910125    &  212          &    X15.7     & 6471   & 2210   &  S16E78  & 5.03  \\
16 & 19910304    &   76          &    X10.4     & 6538   &   -    &  S20E88  & 5.13  \\
17 & 19910322    &  248          &    X13.8     & 6555   & 2530   &  S26E28  & 5.18  \\
18 & 19910601    &  234          &    X28.3     & 6659   & 2300   &  N25E90  & 5.38  \\
19 & 19910604    &  237          &    X32.0     & 6659   & 2300   &  N30E70  & 5.38  \\
20 & 19910606    &  221          &    X30.2     & 6659   & 2300   &  N33E44  & 5.39  \\
21 & 19910609    &  318          &    X15.1     & 6659   & 2300   &  N34E04  & 5.40  \\
22 & 19910611    &  331          &    X17.0     & 6659   & 2300   &  N31W17  & 5.40  \\
23 & 19910615    &  204          &    X34.7     & 6659   & 2300   &  N33W69  & 5.41  \\
24 & 19921102    &  124          &    X13.3     & 7321   & 1650   &  S26W87  & 6.80  \\
25 & 19971106    &  134          &    X13.1     & 8100   & 1000   &  S18W63  & 1.47  \\
26 & 20010402    &  249          &    X29.6     & 9393   & 2440   &  N19W72  & 4.87  \\
27 & 20010415    &  113          &    X21.1     & 9415   & 880    &  S20W85  & 4.91  \\
28 & 20031028    &  247          &    X25.7     & 10486  & 2610   &  S16E08  & 7.44  \\
29 & 20031029    &  250          &    X15.5     & 10486  & 2610   &  S15W02  & 7.45  \\
30 & 20031102    &  138          &    X13.3     & 10486  & 2610   &  S14W56  & 7.46  \\
31 & 20031104    &  64           &    X43.2     & 10486  & 2610   &  S19W83  & 7.46  \\
32 & 20050120    &  65           &    X10.2     & 10720  & 1630   &  N14W61  & 8.68  \\
33 & 20050907    &  24           &    X24.6     & 10808  & 1430   &  S11E77  & 9.30  \\
34 & 20061205    &  46           &    X13.1     & 10930  &  680   &  S07E68  & 10.55 \\
35 & 20110809    &  61           &    X10.7     & 11263  & 1320   &  N17W69  & 2.82  \\
36 & 20170906    &  100          &    X14.8     & 12673  & 2060   &  S08W33  & 8.90  \\
37 & 20170910    &  40           &    X13.0     & 12673  & 2060   &  S08W88  & 8.91  \\
   \enddata
\tablecomments{AR: the NOAA number of S-flare active region, $S_{AR}$: the maximum area of S-flare active region with unit of $10^{-6}$ solar hemisphere ($\mu H$), $t$: the occurring time after the start of the solar cycle. }
\end{deluxetable}

Based on carefully scrutinizing the distribution of S-flares, we find that they can be classified into two types: isolated S-flare and S-flare groups.

(1) Isolated S-flare, refers to only one S-flare generated from an active region. For example, active region NOAA 10808 only generated an X24.6 class flare on 2005 September 6, no other S-flare. Among the 37 S-flares listed in Table 1, there are 23 isolated S-flares (62\%).

(2) S-flare group, refers to at least two or more S-flares generated from an active region, which can be regarded as a group of homologous flares. The most famous S-flare group \textbf{occurred} from October 28 to November 4, 2003 in active region NOAA10486 with 4 S-flares (X25.7, X15.5, X13.3, and X43.2). The most spectacular S-flare group occurred from June 1 to 15, 1991 in active region NOAA6659 with 6 S-flares (from X15.1 to X34.7).

From Figure 1 and Table 1, we may get the following conclusions:

(1) Each solar cycle has only one S-flare group active region, which is NOAA3763 in SC 21 (2 S-flares), NOAA6659 in SC 22 (6 S-flares), NOAA10486 in SC 23 (4 S-flares), and NOAA12673 in SC 24 (2 S-flares). All S-flare groups occurred in the descending phase of solar cycles.

(2) Most S-flares tend to occur after the peaks of solar cycles, $t_{sm}>L_{a}$. This result confirms an earlier conclusion that the more powerful flares tends to occur later in solar cycle (Tan 2011). Among the 37 S-flares, there are 29 (about 78.4\%) occurred in the descending phases of solar cycles, and only 8 (21.6\%) S-flares occurred in the ascending phase. Additionally, except for SC 22 which most S-flares occurred near the peak phase, almost all other S-flares occurred in the descending phase, even far away from the peak point. For example, in SC 21, only one S-flare occurred in the ascending phase (X45.9, on 1978 July 11), all other 8 S-flares occurred in the descending phase, and the last S-flares even occurred on May 20, 1984, 8.07 years after the start of the cycle. In SC 23, the last S-flare even occurred on December 5, 2006, 10.55 years after the start of the cycle.

(3) Most S-flares tend to occur in low-latitude regions on the Sun. Among the 37 S-flares, 26 (70\%) occurred in low-latitude regions with latitude $\leq20$ degrees, while the other 11 S-flares (30\%) occurred in high-latitude regions with latitudes $>$25 degrees. As for the S-flare active regions, 23 of them (85\%) \textbf{occurred} in low-latitude regions and only 4 (15\%) occurred in high-latitude regions. Additionally, all the high-latitude S-flares occurred in SC 22 (from 1989 March to 1991 June).

(4) The occurrence time of the strongest flare ($t_{stg}$) in each solar cycle is strongly anti-correlated with the intensity of cycle ($M_{sn}$, $P_{m}$, and $V_{m}$). The stronger the solar cycle, the earlier the strongest flare occur, while the weaker the cycle, the later the strongest flare occur.

Table 2 lists all parameters of S-flares ($N_{s}$, $t_{s1}$, and $t_{sm}$) and solar cycles ($P_{c}$, $L_{a}$, $L_{d}$, $M_{sn}$, $P_{m}$, $P_{d}$, $V_{m}$, and $V_{d}$) during SC 21 -25. In order to compare different class of flares, we also listed the parameters of X-class flares ($>$X1.0), the extremely strong flares ($>$X14.3, ES-flares), and the strongest flare in each solar cycle. Here, totally there are 719 X-class flares, 37 S-flare, 23 ES-flares during SC 21-24. The magnitudes of all flares occurred before 2017 are corrected by the new calibration of Hudson et al. (2024) with a factor of 1.43.

\begin{deluxetable}{ccccccccccccccccc}
\tablecolumns{16} \tabletypesize{\scriptsize} \tablewidth{0pc}
\tablecaption{The characteristics of daily sunspot number (DSN), X-class flares ($>$X1.0), S-flares ($>$X10.0), ES-flares ($>$X14.3), and the strongest flare in each solar cycle. Here, the magnitudes of all flares occurred before 2017 are corrected by the new calibration of Hudson et al. (2024) with a factor of 1.43.\label{tbl-1}}
\tablehead{\colhead{parameter} &\colhead{~SC21~} &\colhead{~SC22~}  &\colhead{~SC23~} &\colhead{~SC24~} &\colhead{~SC25~} }
  \startdata
 $P_{c}$        & 10.13  &  9.93   & 12.40    &  10.83  &  --           \\
 $L_{a}$        & 4.38   &  4.09   & 4.74     &  4.38   &  --           \\
 $L_{d}$        & 5.75   &  5.84   & 7.66     &  6.45   &  --           \\
 $M_{sn}$       & 428    &  410    & 353      &  220    &  $\geq$290    \\
 $P_{m}$        & 214.1  &  192.2  & 152.1    &  84.7   &  ---            \\
 $P_{d}$        & 61.8   &  64.4   & 51.0     & 33.8    &  ---            \\
 $V_{m}$        & 16.8   &  14.6   & 10.6     &  2.5    &  3.4            \\
 $V_{d}$        & 15.3   &  17.6   & 11.3     &  5.1    &  7.1            \\
 $N_{x}$        & 243    &  205    & 167      &  71     & (108$\pm$5)     \\
 $t_{x1}$       & 0.02   &  1.56   & 0.14     &  1.34   & 1.89            \\
 $t_{xm}$       & 4.75   &  4.00   & 6.05     &  5.12   & --              \\
 $N_{s}$        & 9      &  15     & 10       &  3      & (5$\pm$2)       \\
 $t_{s1}$       & 2.22   &  3.14   & 1.47     &  2.82   & --              \\
 $t_{sm}$       & 6.06   &  4.87   & 6.96     &  6.88   & --              \\
 $N_{es}$       & 6      &  10     & 6        &  1      & (2$\pm1$)       \\
 $t_{es1}$      & 2.22   &  3.14   & 4.87     &  8.90   & ($7.46\pm1.13$) \\
 $t_{esm}$      & 6.28   &  4.79   & 6.91     &  8.90   & ($8.31\pm0.41$) \\
 $t_{stg}$      & 2.22   &  5.41   & 7.46     &  8.90   & ($8.78\pm1.12$) \\
   \enddata
 \tablecomments{$t_{x1}$, $N_{x}$, and $t_{xm}$ are the time of the first X-class flare, the number and the mean time of X-class flares in a solar cycle, respectively. $t_{es1}$, $N_{es}$, and $t_{esm}$ are the time of the first ES-flare, the number and the mean time of ES-flares in a cycle, respectively. $t_{stg}$ is the time of the strongest flare in a solar cycle. The values in parentheses represent the predicted values for SC 25 obtained in this work.}
\end{deluxetable}

Table 3 presents the correlation coefficients (Cr) between the parameters of X-class flares, S-flares, ES-flares and the parameters of solar cycles.

\begin{deluxetable*}{cccccccccccccc}
\tablecolumns{16} \tabletypesize{\scriptsize} \tablewidth{0pc}
\tablecaption{The correlation coefficients (Cr) between parameters of X-class flare ($>$X1.0), S-flare ($>$X10.0), ES-flare ($>$X14.3), the strongest flare and parameters of solar cycles 21 - 24. \label{tbl-1}}
\tablehead{\colhead{Parameter} &\colhead{~~~$t_{x1}$~~~} &\colhead{~~~$t_{xm}$~~~} &\colhead{~~~$N_{x}$~~~} &\colhead{~~~$t_{s1}$~~~} &\colhead{~~~$t_{sm}$~~~} &\colhead{~~~$N_{s}$~~~} &\colhead{~~~$t_{es1}$~~~} &\colhead{~~~$t_{esm}$~~~} &\colhead{~~~$N_{es}$~~~} &\colhead{~~~$t_{stg}$~~~}}
  \startdata
  $P_{c}$     & -0.44  &  0.95 & -0.37  & -0.81  &  0.76 & -0.23  &  0.34 &  0.41  &  -0.29 &  0.56  \\
  $L_{a}$     & -0.72  &  0.98 & -0.23  & -0.94  &  0.85 & -0.36  &  0.22 &  0.45  &  -0.39 &  0.30  \\
  $L_{d}$     & -0.32  &  0.91 & -0.41  & -0.75  &  0.71 & -0.19  &  0.37 &  0.38  &  -0.25 &  0.62  \\
  $M_{sn}$    & -0.39  & -0.38 &  0.99  & -0.13  & -0.64 &  0.82  & -1.00 &  0.90  &   0.86 &  -1.00  \\
  $P_{m}$     & -0.39  & -0.43 &  1.00  & -0.08  & -0.65 &  0.76  & -1.00 &  0.87  &   0.81 &  -1.00  \\
  $P_{d}$     & -0.22  & -0.51 &  0.97  &  0.02  & -0.76 &  0.89  & -0.98 & -0.96  &   0.92 &  -0.82 \\
  $V_{m}$     & -0.40  & -0.40 &  1.00  & -0.11  & -0.64 &  0.77  & \textbf{-1.00} & -0.87  &   0.82  &  \textbf{-1.00} \\
  $V_{d}$     & -0.12  & -0.57 &  0.94  &  0.11  & -0.82 &  0.91  & \textbf{-0.95} & \textbf{-0.98}  &   \textbf{0.95}  &  -0.77 \\
   \enddata
\end{deluxetable*}

Table 3 shows that the time of the first S-flare ($t_{s1}$) and the mean time of all S-flares ($t_{sm}$) in each solar cycle are not correlated with the mean DSN ($V_{m}$) and root-mean-square deviation of DSN ($V_{d}$) in the cycle's valley phase. A similar regime also occurs in X-class flares. These facts indicate that the occurrence of S-flares as well as X-class flares has a clear randomness, and it is difficult to use the parameters of solar cycle to predict their occurrence in advance. Then, what about more powerful flares, the ES-flares?

Here, we choose to define a flare exceeding X14.3 (equal to $>X10$ in old GOES SXR level before Hudson's re-calibration) as an extremely strong flare (ES-flare). Totally, there are 23 ES-flares since 1975, and SC 24 has only one ES-flare, which is enough rare and extremely strong. Table 3 presented the values of Cr between the parameters of ES-flares and solar cycles. We find:

(1) Similar to S-flares, most ES-flares tend to occur after the peak of cycles. Among the 23 ES-flares, 19 (82.6\%) occurred in the descending phases, and only 4 (17.4\%) S-flares occurred in the ascending phase.

(2) Also similar to S-flares, Most ES-flares tend to occur in low-latitude regions on the Sun. Among the 23 ES-flares, 15 (65\%) occurred in low-latitude regions with latitude $\leq20$ degrees, while the other 8 ES-flares (35\%) occurred in high-latitude regions with latitudes $>$25 degrees. As for the ES-flare active regions, 13 of them (81\%) distributed in low-latitude regions and only 3 (19\%) distributed in high-latitude regions.

(3) Different from S-flares and X-class flares, there are strong correlations between the occurrence of ES-flares and the parameters of solar cycles. $N_{es}$ is strongly correlated with $V_{d}$ (Cr=0.95). $t_{es1}$ is strongly anti-correlated with $V_{d}$ (Cr=-0.95) and $V_{m}$ (Cr=-1.00). $t_{esm}$ is strongly anti-correlated with $V_{d}$ (Cr=-0.98). Since the accurate values of $V_{d}$ and $V_{m}$ can be obtained before the start of the solar cycle, it is possible to apply the above strong correlations to predict the occurrence of ES-flares in advance. This fact hints that an essential difference between X-class flares $<$ X14.3 and ES flares $>$ X14.3.

\subsection{Regression, fitting functions, and predictions}

\subsubsection{The number of ES-flares in a solar cycle}

The strong correlation between $N_{es}$ and $V_{d}$ indicates that we may predict $N_{es}$ for the upcoming solar cycle. As $N_{es}$ is an integer rather than real-valued variable, here we apply the Poisson regression (Cameron \& Trivedi 1998, Long \& Freese 2006) to show quantitatively the dependence of $N_{es}$ on $V_{d}$ and $V_{m}$. Figure 2 presents the results of Poisson regression, which predicts that SC 25 possibly will have 2$\pm$1 (or say: 1$\sim$3) ES-flares. Here, the regression between $N_{es}$ and $V_{d}$ output the prediction is 2.3$\pm$1.1, while the regression between $N_{es}$ and $V_{m}$ output the prediction is 2.2$\pm$1.2. Both results are close to each other, but the former has higher credibility for its Pseudo R-squ. is higher (0.8372) than the latter (0.7466). The left panel of Figure 2 shows that the observational results distributed approximately along the diagonal, while right panel shows that the results distributed more disperse from the diagonal.

\begin{figure} 
\begin{center}
   \includegraphics[width=8.4 cm]{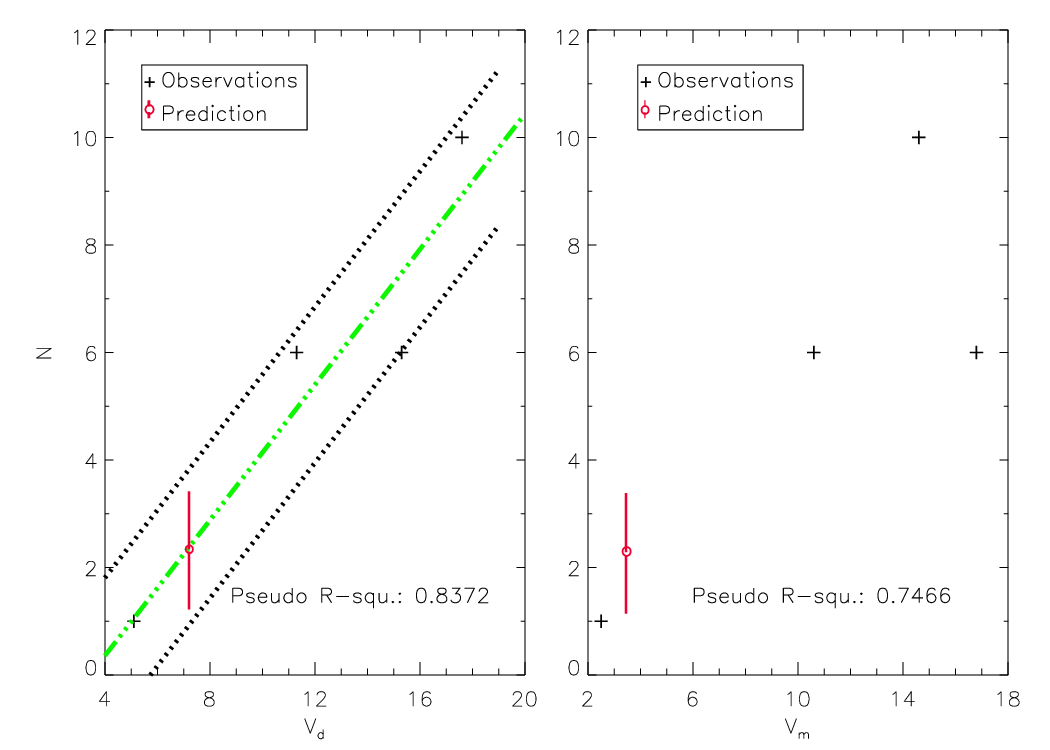}
  \caption{The results of Poisson regression of the number of ES-flares ($N_{es}$) in solar cycles vs the root-mean-square deviation of DSN ($V_{d}$, left) and mean DSN ($V_{m}$, right) around the cycle valley. The green thick dash-dotted line is a heuristic fitting function obtained from linear regression, the red circle is the predicting results for SC 25 in this work.}
\end{center}
\end{figure}

Because of the high correlation coefficient between $N_{es}$ and $V_{d}$ (Cr$=0.95$), as a comparison, we also apply a linear regression to show a heuristic fitting function (overplotted in green thick dash-dotted line in the left panel of Figure 2), $N_{es}\approx -2.16+0.63V_{d}\pm1.05$. As for SC 25, $V_{d}=$ 7.1 and $V_{m}=$ 3.4. Substituting these values into above expression, then $N_{es}\approx2.31\pm1.05$, which also close to the result derived from the above Poisson regression, or simply to say, SC 25 will have 1$\sim$3 ES-flares. Using the same method and the strong correlations between $N_{x}$ and $V_{m}$ (Cr$=1.00$), and between $N_{s}$ and $V_{d}$ (Cr$=0.91$), we derived that SC 25 will have 108$\pm$5 X-class flares and 5$\pm$2 S-flares.

The above results imply that the number of ES-flares in the upcoming solar cycle is strongly depending on $V_{d}$. The higher the $V_{d}$, the more the ES-flares occurring in that cycle. The left panel of Figure 2 showed that the distribution of SC 21 - 24 are approximately concentrated near the diagonal.

\subsubsection{The occurrence time of ES-flares}

Then, when will these ES-flares occur? Let's first examine the mean time of occurrence of ES-flares ($t_{esm}$) in each cycle. Table 2 shows that the strongest correlation takes place between $t_{esm}$ and $V_{d}$ with Cr=-0.98. The red triangles of left panel of Figure 3 plots the distribution which concentrated approximately along the opposite diagonal. From the linear regression, we may obtain a best fitting function,

\begin{equation}
t_{esm}\approx 10.48-0.31V_{d}\pm0.30.
\end{equation}

Substituting the value of $V_{d}$ for SC 25, we obtain $t_{esm}\approx8.31\pm0.30$ yr. This implies that the possible 2$\pm$1 ES-flares in SC 25 will take place averagely 8.31 years after the valley point (2019 August 11).

\begin{figure} 
\begin{center}
   \includegraphics[width=8.4 cm]{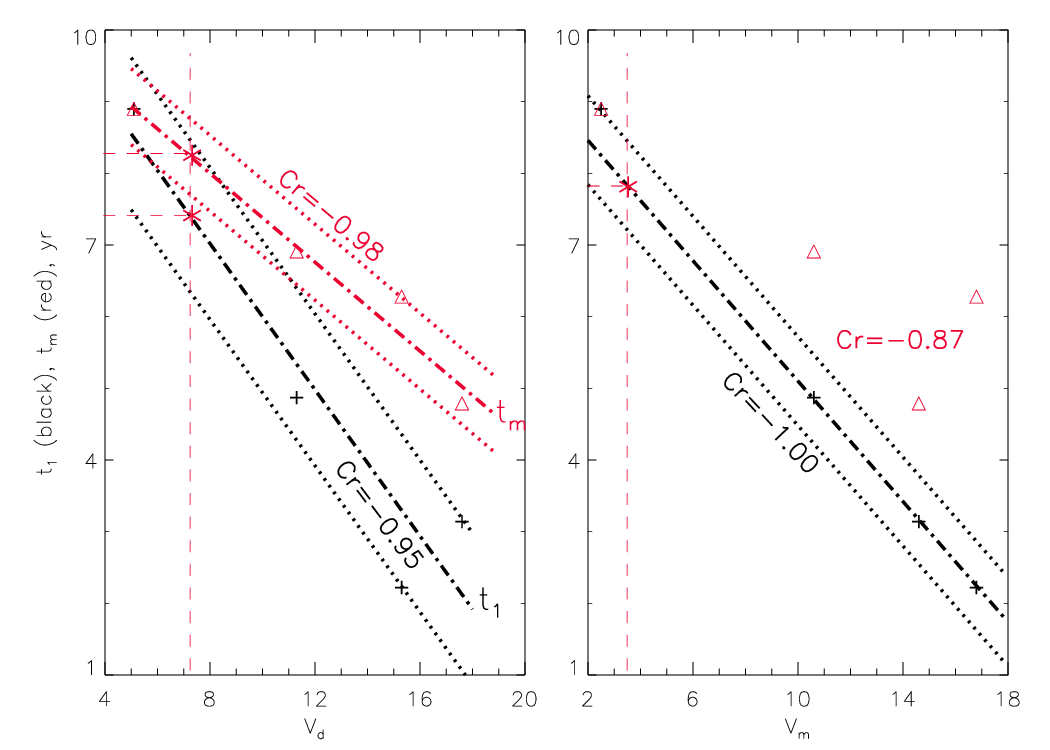}
  \caption{The occurring time of ES-flares in solar cycles vs the root-mean-square deviation of DSN ($V_{d}$, left) and mean DSN ($V_{m}$, right) during the cycle valley. The black parts (including +, the fitted thick dash-dotted lines, and thin dotted lines ) show the time of the first ES-flare in solar cycles $t_{es1}$, while red parts show the mean time of all ES-flares in solar cycle, $t_{esm}$. The red $*$ is the predicting results for SC 25 in this work.}
\end{center}
\end{figure}

Now, we try to answer another question: when will the first ES-flare occur in a solar cycle? Table 2 shows that the correlation coefficient between $t_{es1}$ and $V_{m}$ is high up to -1.00, a very strong anti-correlation, and the related distribution is shown in the right panel of Figure 3 (black part). The linear regression derived a best fitting function,

\begin{equation}
t_{es1}\approx 9.33-0.47V_{m}\pm0.69.
\end{equation}

With this function, we may get a prediction for SC 25: $t_{es1}\approx7.73\pm0.69$ yr. That is to say, the first ES-flare will occur around May of 2027. At the same time, the correlation coefficients between $t_{es1}$ and $V_{d}$ is -0.95, also a strong anti-correlation, and the related distribution is shown in the left panel of Figure 3 (black), and the linear regression function can be obtained,

\begin{equation}
t_{es1}\approx 11.10-0.51V_{d}\pm0.80.
\end{equation}

Substituting $V_{d}=$ 7.10 into the above expression, we get $t_{es1}=7.48\pm0.80$ yr, it is about in January or February of 2027. Additionally, we may attempt to apply binary linear regression to obtain a best fitting function of $t_{es1}$ with respect to $V_{d}$ and $V_{m}$,

\begin{equation}
t_{es1}\approx 9.45-0.46V_{m}-0.0125V_{d}\pm0.62.
\end{equation}

It derived: $t_{es1}\approx7.80\pm0.62$ yr (around May or June of 2027). Obviously, the above three predicting results ($t_{es1}=7.73\pm0.69$, $7.48\pm0.80$, and $7.80\pm0.62$ yr) are essentially very close to each other. In order to ensure a more reliable prediction, we list the predicted value of $7.48\pm0.80$ for SC 25 in Table 2. In brief, SC 25 will possibly have about 1 - 3 ES-flares, and all of them will occur after the spring of 2027.

\subsubsection{The occurrence time of the strongest flare}

Using the strong anti-correlation between $t_{stg}$ and $V_{m}$ ($Cr=-1.00$), we may obtain the linear regression function,

\begin{equation}
t_{stg}\approx 10.11-0.393V_{m}\pm0.93.
\end{equation}

It derived that the occurrence time of the strongest flare in SC 25 is about $t_{stg}\approx 8.78\pm0.93$ yr (around May of 2028).

Here, it is necessary to emphasize that the above results are only based on small dataset with 4 solar cycles since 1975, there should contain significant uncertainty. Strictly speaking, it is difficult to obtain exact statistical results with so small dataset. Therefore, we chose the parameter pairs with high correlation coefficient $\geq$0.95 for a heuristic prediction, and the corresponding confidence level exceeds 95\%. Table 3 listed that there are other three pairs of parameters between $t_{es1}$ and $M_{sn}$, $P_{m}$, $P_{d}$ during the peak phase are anti-correlated with coefficients of Cr exceeding 0.95. Because $M_{sn}$, $P_{m}$ and $P_{d}$ actually reflect the intensity of solar cycle, these anti-correlations imply that the stronger the solar cycle, the earlier ES-flares occur. However, due to most ES-flares occurred after the peak phase or even in the descending phase of solar cycles, the application of predicting ES-flares has little practical significance. On the contrary, the valley parameters of solar cycles are generally at least 2 years earlier than the first ES-flares (last column in Table 1), and they can serve as precursors for predicting ES-flares, which has practical operability.

\subsection{Physical explanations}

Why do most of the S-flares and ES-flares occurred in the late phase of solar cycles and low-latitude regions on the solar disk? Why is the occurrence of S-flares in the solar cycle somewhat accidental, while the occurrence of ES-flares is significantly correlated with DSN during the valley phase of solar cycles?

Generally, large active regions should have more complex magnetic field structures and are more prone to producing strong flares (Sammis et al. 2000), such as the super active region (SAR) with $\beta\gamma\delta$-tpye magnetic configuration (Bai 1987, Tian et al. 2002, Chen et al. 2011). Table 1 listed the NOAA number, maximum area and location of ES-flare active regions (ESARs), which shows that the 23 ES-flares originated from 16 ESARs, of which 12 have a maximum area exceeding 1000$\mu H$ and should be super active regions (SARs). At the same time, we also noted that many very large $\beta\gamma\delta$-type complex SARs had no S-flare or ES-flare, such as NOAA3776 (3100$\mu H$), 5669 (3080$\mu H$), 12192 (2710$\mu H$), etc. These facts indicate that SARs are not the essential condition for generating ES-flares. The generation of ES-flares should be related to the early evolution of solar cycles, such as $V_{m}$ and $V_{d}$.

Let's look at the latitude distribution of valley sunspot regions (VSRs). Figure 4 presents their latitude distributions in SC 23, SC 24, and SC25. Most of them are simple $\alpha$ or $\beta$ type active regions, with area of less than 150 $\mu H$ and lifetime of 2 - 10 days. There are about 75\% of VSRs distributing in a zone within 12 degrees north and south latitude of the Sun, while the other 25\% of VARs are distributed in high latitudes beyond 12 degrees to about 40 degrees. This fact indicates that even during the valley phases of solar cycles, although the sunspot number is very small, magnetic activity still exists in the low-latitude region. It hints that the low-latitude region may be a long-term active zone possibly different from the new emerging sunspot region in the upcoming solar cycle, called as low-latitude active zone (LAZ), which always keeps a certain level of activity during the valley phase of the solar cycle. Solar dynamo theory indicated that the toroidal magnetic field generated from a poloidal field due to the solar differential rotation. In each solar cycle, the new emerging sunspot region first appeared in high-latitude regions (near $40^{\circ}$) and then gradually migrate to low-latitude regions, and finally arrive at about $5-8^{\circ}$ latitude near the end of the cycle (Babcock 1961, Leighton 1969). Due to the enormous energy accumulation required for ESAR to generate ES-flares, it is not easy to provide such conditions in a typical active region. Considering that most ES-flares occurred in regions with latitudes below 20 degrees and in the late phase of solar cycle, we suggest that ES-flares should be mainly generated from the interaction between the newly emerging active region and LAZ. Such interaction may lead to large-scale energy accumulation and most powerful eruption. In the early phase (such as the ascending phase) of solar cycles, the newly emerging active regions located in high latitude and far from LAZ, their interaction should be very weak, and difficult to produce ES-flares. On the contrary, in the late phase (such as the descending phase) of solar cycles, the newly emerging active regions occurred in low latitude and very close to LAZ, their interaction should become very strong and tend to produce extremely powerful ES-flares. The values of $V_{d}$ and $V_{m}$ are mainly dominated by the sunspots in LAZ during the valley phase before the onset of solar cycles. Tan (2019) reported that the strength of a solar cycle is related to the mean DSN during the valley phase. Table 2 and 3 also show that, besides being strongly anti-correlated with $V_{d}$ and $V_{m}$, $t_{es1}$ is also strongly anti-correlated with $M_{sn}$, $P_{m}$, and $P_{d}$. This actually indicates that the stronger the solar cycle, the more the ES-flares, and the earlier ES-flares occurred.

\begin{figure} 
\begin{center}
    \includegraphics[width=8.4 cm]{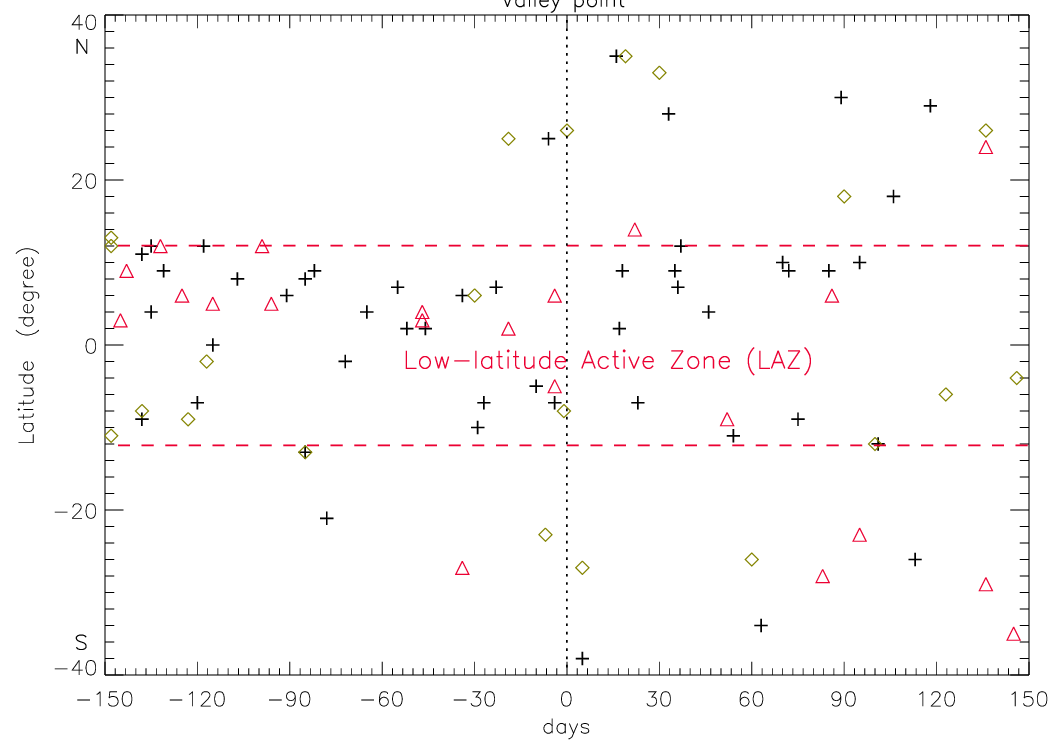}
  \caption{The latitude distribution of sunspot active regions occurred during valley phases of SC 23 (black +, from 1995 December 20 to 1996 October 15), SC24 (green $\diamond$, from 2008 May 14 to 2009 March 11), and SC25 (red $\triangle$, from 2019 March 14 to 2020 January 8). The 0 point on the X-axis represents the valley point between two solar cycles. The area between two red dashed lines represents the low-latitude active zone (LAZ).}
\end{center}
\end{figure}

As mentioned in Section 3.1 and Table 1, there are 8 ES-flares occurred in high-latitude regions beyond 25 degrees. All of them generated from 3 ESARs (NOAA5395, 5747, and 6659) around the peak phase of SC 22 (from March 1989 to June 1991). Especially, the active region NOAA6659 produced 6 ES-flares, demonstrating the uniqueness of this solar cycle. This fact indicates that in addition to the interaction between LAZ and the newly emerging active regions, other mechanisms may also contribute to generate ES-flares, such as interactions between the nearby active regions (collisions, mergers, and magnetic induction, etc.) or the evolution of large active regions themselves. Among the 23 ES-flares, 8 come from high-latitude regions, accounting for 35\%; and among the 16 ESARs, there are 3 high-latitude ESARs, accounting for 19\%. This indicates that the interaction between LAZ and the newly emerging active regions should be the main mechanism for the formation of ES-flares.

\section{Summary and Discussion}

Based on the analysis of the daily sunspot numbers and GOES SXR observations since 1975, we obtained the following conclusions regarding to S-flares:

(1) Both S-flares and the stronger ES-flares mostly occur in the late phases of solar cycles and low-latitude regions on the solar disk.

(2) Similar to X-class flares, the of S-flares ($>$X10) in each solar cycle is somewhat random, but ES-flares ($>$X14.3) seem to be dominated by $V_{m}$ and $V_{d}$. The number of ES-flares in a cycle is strongly correlated with $V_{d}$, and the time of the first ES-flare is strongly anti-correlated with $V_{m}$ and $V_{d}$. The mean time of all ES-flares in a cycle is strongly anti-correlated with $V_{d}$ (Figure 2 and 3). The stronger the solar cycle, the earlier the strongest flare occurs.

(3) $V_{d}$ and $V_{m}$ can be applied to predict the number and timing of ES-flares in the forthcoming solar cycles. The linear regressions may derive some heuristic fitting functions of $N_{es}$ vs $V_{d}$, $t_{es1}$ vs $V_{m}$, $t_{es1}$ vs $V_{d}$, and $t_{esm}$ vs $V_{d}$, and $t_{stg}$ vs $V_{m}$. Based on these strong correlations and regression fitting functions, we may predict the ES-flares in SC 25: about $2\pm1$ ES-flares, and the first ES-flare may occur around the spring of 2027, which is slightly close to the end of the cycle. This result indicates that for space weather forecasting, 2027 possibly should be a key year and require special attentions.

Here, it should be emphasized that the above conclusions \textbf{are} based on a small dataset of the past 4 solar cycles, the above conclusions is only heuristic and requires further verification with more observations in the future.

Further analysis shows that the occurrence of ES-flares during a solar cycle is closely related to the cycle intensity: the stronger the solar cycle, the higher the $V_{m}$ and $V_{d}$, the more ES-flares there are, and the earlier they occur. We proposed that the Sun may have a LAZ, and most ES-flares are the results of the interaction between the newly emerging active regions and LAZ. In the early phase of solar cycle, the newly emerging active region is located in the high-latitude region, far away from LAZ, with weak interaction, making it less likely to produce ES-flares. On the contrary, in the late phase of solar cycle, the newly emerging active region is located in the low-latitude region, close to LAZ, with strong interaction and easy to generate ES-flares. This explains why most of the ES-flares occurred in the late phase of solar cycles and near the low-latitude regions on the solar disk.

Here, we are also soberly aware that LAZ is only a preliminary assumption just derived from the limited observations, and requires more observations and theoretical researches to confirm. In fact, LAZ is somewhat similar to the equatorial low-pressure zone formed by the compression of trade winds from the northern and southern hemispheres of Earth, mainly due to the retation of celestial body. The rotation of the Sun is also likely to form a low-latitude active zone (LAZ). In the near future, we will continue to delve into the characteristics and evolutions of the solar LAZ and the generations of ES-flares, in order to provide more reliable physical basis for understanding the generation of ES-flares, predicting their occurrence and the possible disastrous space weather events.

\begin{acknowledgments}
This work is supported by the Strategic Priority Research Program of the Chinese Academy of Sciences XDB0560000, the National Key R\&D Program of China 2021YFA1600503, 2022YFF0503001, and the International Partnership Program of Chinese Academy of Sciences 183311KYSB20200003.
\end{acknowledgments}


\begin{thebibliography}{}
\bibitem[Babcock(1961)]{Babcock1961}Babcock, H.W.: 1961, \emph{ApJ}, \textbf{133}, 572.
\bibitem[Bai(1987)]{Bai1987}Bai, T.: 1987, \emph{ApJ}, \textbf{314}, 795.
\bibitem[Bai(1988)]{Bai1988}Bai, T.: 1988, \emph{ApJ}, \textbf{328}, 860.
\bibitem[Bai(1989)]{Bai1989}Bai, T., Sturrock, P.A.: 1989, \emph{Ann. Rev. Astron. Astrophys.}, \textbf{27}, 421.
\bibitem[Bloomfield(2012)]{Bloomfield2012}Bloomfield, D. S., Higgins, P.A., McAteer, R.T.J., Gallagher, P.T.: 2012, \emph{ApJL}, \textbf{747}, L41.
\bibitem[Cameron(1998)]{Cameron1998}Cameron, A. C. and Trivedi, P. K.: 1998, Regression Analysis of Count Data. New York: Cambridge Press.
\bibitem[Chen(2011)]{Chen2011}Chen, A.Q., Wang, J.X., Li, J.W., Feynman, J., Zhang, J.: 2011, \emph{Astron. Astrophys.}, \textbf{534}, A47.
\bibitem[Clette(2014)]{Clette2014}Clette, F., Leif, S., Jose, M.V., Edward, W. C.: 2014, \emph{Space Sci Rev}, \textbf{186}, 35.
\bibitem[Clette(2016)]{Clette2016}Clette, F., Lefevre, L.: 2016, \emph{Sol Phys}, \textbf{291}, 2629.
\bibitem[Cliver(2022)]{Cliver2022}Cliver, E. W., Schrijver, C.J., Shibata, K., Usoskin, I.G.: 2022, \emph{Living Rev. Sol Phys}, \textbf{19}, 2.
\bibitem[Emslie(2012)]{Emslie2012}Emslie, A. G., Dennis, B.R., Shih, A.Y., Chamberlin, P.C., Mewaldt, R.A., Moore, C.S., et al.: 2012, \emph{ApJ}, \textbf{759}, 71.
\bibitem[Guo(2021)]{Guo2021}Guo, W., Jiang, J., Wang J.X.: 2021, \emph{Sol Phys}, \textbf{296}, 136.
\bibitem[Hale(1908)]{Hale1908}Hale, G. E.: 1908, \emph{ApJ}, \textbf{28}, 315.
\bibitem[Hayakawa(2023)]{Hayakawa2023}Hayakawa, H., Bechet, S., Clette, F., Hudson, H.S., Maehara, H., Namekata, K., Notsu, Y.: 2023, \emph{ApJ}, \textbf{954}, L3.
\bibitem[Harrison(1995)]{Harrison1995}Harrison, R.A.: 1995, \emph{Astron. Astrophys.}, \textbf{304}, 585.
\bibitem[Hudson(2007)]{Hudson2007}Hudson, H.S.: 2007, \emph{ApJ}, \textbf{663}, L45.
\bibitem[Hudson(2021)]{Hudson2021}Hudson, H.S.: 2021, \emph{Annu. Rev. Astron. Astrophys.}, \textbf{59}, 445.
\bibitem[Hudson(2024)]{Hudson2024}Hudson, H.S., Cliver, Ed, White, S., Machol, J., Peck, C., Tolbert, K., Viereck, R., Zarro, D.: 2024, \emph{Sol Phys}, \textbf{299}, 39.
\bibitem[Jiang(2023)]{Jiang2023}Jiang, J., Zhang Z.B., Petrovay K.: 2023, \emph{JASTP}, \textbf{234}, 106018.
\bibitem[Kane(2005)]{Kane2005}Kane, S.R., McTiernan, J.M., Hurley, K: 2005, \emph{Astron. Astrophys.}, \textbf{433}, 1133.
\bibitem[Khlystov(2014)]{Khlystov2014}Khlystov, A.I.: 2014, \emph{Atmospheric and Oceanic Phys.}, \textbf{50}, 776.
\bibitem[Le(2021)]{Le2021}Le, G. M., Liu, G. A., Zhao, M.X., Mao, T., Xu, P.G.: 2021, \emph{RAA}, \textbf{21}, 130.
\bibitem[Leighton(1969)]{Leighton1969}Leighton, R. B.: 1969, \emph{ApJ}, \textbf{156}, 1.
\bibitem[Li(2002)]{Li2002}Li, K.J., Zhan, L.S., Wang, J.X., Liu, X.H., Yun, H.S., Xiong, S.Y., et al: 2002, \emph{Astron. Astrophys.}, \textbf{392}, 301.
\bibitem[Li(2023)]{Li2023}Li, D., Warmuth, A., Wang, J.C., Zhao, H.S., Lu, L., Zhang, Q.M., et al: 2023, \emph{RAA}, \textbf{23}, 095017.
\bibitem[Long(2006)]{Long2006}Long, J. S. and Freese, J.: 2006, Regression Models for Categorical Dependent Variables Using Stata. Second Edition. College Station, TX: Stata.
\bibitem[Luo(2024)]{Luo2024}Luo, P.X, Tan, B.L.: 2024, \emph{RAA}, \textbf{24}, 035016.
\bibitem[Maehara(2012)]{Maehara2012}Maehara, H., Shibayama, T., Notsu, S., Notsu, Y., Nagao, T., Kusaba, S., Honda, S., Nogami, D, Shibata, K.: 2012, \emph{Nature}, \textbf{485}, 478.
\bibitem[Notsu(2019)]{Notsu2019}Notsu, Y., Maehara, H., Honda, S., Hawley, S.L., Davenport, J.R.A., Namekata, K., et al.: 2019, \emph{ApJ}, \textbf{876}, 58.
\bibitem[Priest(2002)]{Priest2002}Priest, E.R., Forbes, T.G.: 2002, \emph{Astron. Astrophys. Rev.}, \textbf{10}, 313.
\bibitem[Romano(2007)]{Romano2007}Romano, P., Zuccarello, F.: 2007, \emph{Astron. Astrophys.}, \textbf{474}, 633.
\bibitem[Schrijver(2007)]{Schrijver2007}Schrijver, C. J.: 2007, \emph{ApJ}, \textbf{655}, L117.
\bibitem[Sammis(2000)]{Sammis2000}Sammis, I., Tang, F., Zirin, H: 2000, \emph{ApJ}, \textbf{540}, 583.
\bibitem[Schaefer(2000)]{Schaefer2000}Schaefer, B.E., King, J.R., Deliyannis, C.P.: 2000, \emph{ApJ}, \textbf{529}, 1026.
\bibitem[Shibayama(2013)]{Shibayama2013}Shibayama, T., Maehara, H., Notsu, S., Notsu, Y., Nagao, T., Honda, S., Ishii, T.T., Nogami, D, Shibata, K.: 2013, \emph{ApJS}, \textbf{209}, 5.
\bibitem[Tan(2011)]{Tan2011}Tan, B.L.: 2011, \emph{Astrophy. Space Sci.}, \textbf{332}, 65.
\bibitem[Tan(2019)]{Tan2019}Tan, B.L.: 2019, \emph{Adv. Space Res.}, \textbf{63}, 617.
\bibitem[Tian(2002)]{Tian2002}Tian, L.R., Liu, Y., Wang, J.X.: 2002, \emph{Solar Phys.}, \textbf{209}, 361.
\bibitem[Upton(2023)]{Upton2023}Upton, L.A., Hathaway, D.H.: 2023, \emph{JGR - Space Phys.}, \textbf{128}, e2023JA031681.
\bibitem[Zhang(2003)]{Zhang2003}Zhang, H.Q., Bao, X.M., Zhang, Y., Liu, J.H., Bao, S.D., Deng, Y.Y., et al.: 2003, \emph{Chin. J. Astron. Astrophys.}, \textbf{3}, 491.

\end{thebibliography}
\end{document}